\documentclass[aps, pre, showpacs, showkeys, twocolumn]{revtex4-1}

\usepackage{amsmath,amssymb,amsthm,amsfonts,latexsym}
\usepackage{bm,amsfonts, mathtools}
\usepackage{graphicx,color, wasysym}
\usepackage{textcomp}
\usepackage{amsmath,amssymb,latexsym,epsfig}
\usepackage{appendix}
 \usepackage{hyperref}
 \usepackage{color}

\def\ulamek#1#2{\mbox{\normalfont$\frac{#1}{#2}$}}

\DeclareMathOperator{\okr}{{\stackrel{{\scriptscriptstyle{\mathsf{def}}}}{=}}}
\DeclareMathOperator{\D}{d\!}
\DeclareMathOperator{\E}{e}

\begin{document}

\title[Mittag-Leffler function and fractional differential equations]{Mittag-Leffler function and fractional differential equations}

\author{K.~G\'{o}rska} 
\email{katarzyna.gorska@ifj.edu.pl}
%\affiliation{H. Niewodnicza\'{n}ski Institute of Nuclear Physics, Polish Academy of Sciences, Division of Theoretical Physics, ul. Eliasza-Radzikowskiego 152, PL 31-342 Krak\'{o}w, Poland}

\author{A.~Lattanzi}
\email{ambra.lattanzi@gmail.com}
\affiliation{H. Niewodnicza\'{n}ski Institute of Nuclear Physics, Polish Academy of Sciences, Division of Theoretical Physics, ul. Eliasza-Radzikowskiego 152, PL 31-342 Krak\'{o}w, Poland}

\author{G. Dattoli}
\email{giuseppe.dattoli@enea.it}
\affiliation{ENEA - Centro Ricerche Frascati, via E. Fermi, 45, IT 00044 Frascati (Roma), Italy}

\begin{abstract}
We adopt a procedure of operational-umbral type to solve the $(1+1)$-dimensional fractional Fokker-Planck equation in which time fractional derivative of order $\alpha$ ($0 < \alpha < 1$) is in the Riemann-Liouville sense. The technique we propose merges well documented operational methods to solve ordinary FP equation and a redefinition of the time by means of an umbral operator. We show that the proposed method allows significant progress including the handling of operator ordering.
\end{abstract}

\keywords{fractional Fokker-Planck equation, fractional calculus, moments, umbral (operational) method}

%\pacs{..}

\maketitle

\section{Introduction}\label{sec1} %%%%%%%%%%%%%%%%%%%%%%

The fractional Fokker-Planck (fFP) equation is widely used in physics to describe the continuous time random walk connected with the anomalous diffusion \cite{MMagdziarz07, MMagdziarz15, RMetzler00, YuPovstenko17, TSandev16, IMSokolov06}. It has the form
\begin{equation}\label{1.1}
\partial_{t}^{\alpha} F_{\alpha}(x, t) = L_{F\!P} F_{\alpha}(x, t) + \frac{t^{-\alpha}}{\Gamma(1-\alpha)}f(x)
\end{equation}
with $f(x) = F_{\alpha}(x, 0)$ and the fractional time derivative in the Riemann-Liouville sense \cite{AAKilbas06, IPodlubny99} of $\alpha$th order where $0 < \alpha < 1$. We assume $0 < \alpha < 1$, $t \geq 0$, and Cauchy boundary condition this is $x\in\mathbb{R}$. The physicists and mathematicians also considered the fFP equation with the so-called signaling \cite{RGorenflo98, YuLuchko13, FMainardi07} and Dirichlet \cite{YuLuchko12, YZhou14} boundary condition.

The most common approach of solving the fFP equation is presented in, e.g., \cite{AAKilbas06, IPodlubny99}. It bases on using the Laplace and Fourier transforms \cite{INSneddon74} which they applied into Eq. \eqref{1.1} give
\begin{equation}\label{1.2}
F_{\alpha}^{(LF)}(\kappa, \tau) = \frac{\tau^{\alpha-1}}{\tau^{\alpha} - L^{(F)}_{FP}} f^{(F)}(\kappa).
\end{equation}
The Laplace-Fourier transforms is denoted by the superscript $(LF)$ whereas the Fourier transform by the superscript $(F)$. In this paper we assume that the Fourier transform of initial condition $f(x)$ exists. According to the standard procedure $F_{\alpha}(x, t)$ can be obtained by inverting the Laplace and Fourier transforms in Eq. \eqref{1.2}. Indeed, using
\cite[Eq. (1.80)]{IPodlubny99} we can write the solution of fFP equation as
\begin{equation}\label{1.3}
F_{\alpha}(x, t) = \int_{-\infty}^{\infty}\E^{i\kappa x} E_{\alpha}(L_{F\!P}^{(F)} t^{\alpha}) f^{(F)}(\kappa) \D\kappa, \quad 0 < \alpha < 1.
\end{equation}
In Eq. \eqref{1.3} the Mittag-Leffler (ML) function $E_{\alpha}(\sigma)$ is given for $0 < \alpha < 1$ and $\sigma\in\mathbb{R}$. The ML function can be defined by the contour integral where the contour is chosen in such a way which allows us to omit its singularities \cite{RGorenflo14, AAKilbas06, IPodlubny99}. These singularities can generate many problems in numerical calculation of $E_{\alpha}(\sigma)$; some of them are solved in \cite{RGarrappa13}. Note that due to the evolution operator method, which is employed to solve for instance the time dependent Schr\"{o}dinger equation, Eq. \eqref{1.3} can also be interpreted as the action of evolution operator $U_{\alpha}(t) = E_{\alpha}(t^{\alpha}L_{F\!P})$ onto $f(x)$.

In this paper we express $F_{\alpha}(x, t)$ in terms of the solution $F_{1}(x, t)$ of standard (ordinary) Fokker-Planck equation, namely Eq. \eqref{1.1} for $\alpha = 1$. It is obtained by employing the so-called shift (umbra) representation of ML function. The umbral approach is frequently applied in the characterization of special functions  and polynomials \cite{GDattoli17, IMGessel05, SRoman84} especially in the combinatorial description of the latter one \cite{IMGessel05, GCRota73, VStrehl17}. Usually, in many cases the use of umbral operational technique notably simplifies the calculations like, e.g., integrations and summations \cite{DBabusci17, MMDzherbashyan66, SRoman84}. For that reasons we believe that it will happen in the case of this paper and it will help in derivation of the exact and explicit of $F_{\alpha}(x,t)$ and their $n$th moments.

The paper is organized as follows. The three representation of ML function, namely its series, integral and shift (umbra) representations, are described in Section \ref{sec2}. There is also shown the justification of using the shift (umbra) version of ML function. Section \ref{sec3} contains the considerations about the convolution properties of the ML function and, at the same, the convolution property of the evolution operator. Employing the umbral representation of $F_{\alpha}(x, t)$ we will present the solution of fFP equation as the solution of standard (ordinary) FP equation with changed time, see Section \ref{sec4}. In Section \ref{sec5} we will apply the shift (umbral) form of $F_{\alpha}(x,t)$ to derive the $n$th moments for various examples of the FP operator. The paper is concluded in Section \ref{sec6}.

\section{The ML function}\label{sec2} %%%%%%%%%%%%%%%%%%%%%

We begin with the ML functions whose series expansion reads \cite{RGarrappa13, RGorenflo14, AAKilbas06, IPodlubny99}
\begin{equation}\label{2.4}
E_{\alpha}(z) = \sum_{r=0}^{\infty} \frac{z^{r}}{\Gamma(1+\alpha r)}, \quad \alpha>0 \quad \text{and} \quad z\in\mathbb{C}.
\end{equation}
Eq. \eqref{2.4} for $\alpha = 1$ is equal to $\exp(z)$ whereas for $\alpha = 1/2$ it is equal to $\E^{z^{2}}[{\rm erf}(z) + 1]$. The error function is traditional denoted as ${\rm erf}(z)$. For a real argument and $0 < \alpha < 1$ the ML function can be also expressed as
\begin{equation}\label{2.5}
E_{\alpha}(\lambda x) = \int_{0}^{\infty} n_{\alpha}(y, x^{1/\alpha}) \E^{\lambda y} \D y,
\end{equation}
where
\begin{equation}\label{2.6}
n_{\alpha}(y, x) = \frac{x}{\alpha y^{1+1/\alpha}} \varPhi_{\alpha}(x/y^{1/\alpha}),
\end{equation}
see \cite[Eq. (7)]{KWeron96} and \cite[Eq. (11)]{KGorska12}. The one-sided L\'{e}vy stable distribution $\varPhi_{\alpha}(u)$, $0 < \alpha < 1$,  is uniquely defined though the inverse Laplace transform of the stretched exponential $\exp(-p^{\alpha})$  named also as the Kohlrausch-Williams-Watts function or the KWW function  \cite{HPollard46}. For $\alpha = l/k$, $l<k$, it can be presented as the finite sum of $k-1$ generalised hypergeometric function \cite[Eqs. (3) and (4)]{KAPenson10}:
\begin{align}\label{2.7}
\begin{split}
\varPhi_{\frac{l}{k}}(u) & = \sum_{j=1}^{k-1} \frac{(-1)^{j} u^{-1-j\frac{l}{k}}}{j! \Gamma(-j\frac{l}{k})} \\
& \times {_{l+1}F_{k}}\left({1, \Delta(l, 1 + j \frac{l}{k}) \atop \Delta(k, 1+j)}; (-1)^{k-l} \frac{l^{l}}{k^{k} u^{l}}\right).
\end{split}
\end{align}
To get Eq. \eqref{2.7} from \cite[Eqs. (3) and (4)]{KAPenson10} we employed the Euler's reflection formula for gamma functions to \cite[Eq. (4)]{KAPenson10}. The symbol $\Delta(n, a)$ is the sequence of numbers $a/n, (a+1)/n, \ldots, (a+n-1)/n$. Note also that Eq. \eqref{2.7} leads to the series representation of $\varPhi_{l/k}(u)$ presented in  \cite[Eq. (4)]{HPollard46}. This can be shown by writing ${_{l+1}F_{k}}$ as the series \cite[Chapter 7]{APPrudnikov-v3} and applying the splitting formula for obtained sums. Eq. \eqref{2.7} for $k \leq 3$ reduces to the special functions, e.g. for $\alpha=1/2$ it is $\varPhi_{1/2}(u) = \exp[-1/(4u)]/(2\sqrt{\pi} u^{3/2})$. The remain examples are listed in \cite{KAPenson10}.

The one-sided L\'{e}vy stable distribution satisfies the unusual properties like, e.g., essential singularities at $u=0$, 'heavy-long' tail asymptotic behaviour for $u\to \infty$, and the self-similar properties $\varPhi_{\alpha}(x, y) = y^{-1/\alpha} \varPhi_{\alpha}(x/y^{1/\alpha})$. The 'heavy-long' tail of $\varPhi_{\alpha}(u)$ ensures that the Stieltjes moments
\begin{equation}\label{2.8}
M_{\alpha}(\sigma) = \int_{0}^{\infty} u^{\, \sigma}\varPhi_{\alpha}(u) \D u = \frac{\Gamma(1-\sigma/\alpha)}{\Gamma(1-\sigma)}, \quad 0 < \alpha < 1,
\end{equation}
are infinite for $\sigma \geq \alpha$ and finite otherwise. We point up that due to the converse Carleman criterion for non-uniqueness \cite[Criterion {\bf C3}]{AHorzela10} this Stieltjes moment problem is non-unique for $k-l > 3$.

The ML function given by Eq. \eqref{2.4} can also be written as the series which contains the Stieltjes moments $M_{\alpha}(\sigma)$, namely
\begin{equation}\label{2.9}
E_{\alpha}(x) = \sum_{n=0}^{\infty} \frac{x^{n}}{n!} M_{\alpha}(-\alpha n).
\end{equation}
This form of ML function reconstructs Eq. \eqref{2.4} when we substitute the second equality in Eq. \eqref{2.8} into Eq. \eqref{2.9}. The dependence on $n$ in $M_{\alpha}(-\alpha n)$ can be eliminated by introducing the shift operator $c_{\rho}$ (called also the umbral operator) such that $c^{\mu}_{\rho}: \gamma(\rho)\mapsto \gamma(\rho+\mu)$ and $c^{\mu}_{\rho}c^{\nu}_{\rho} = c^{\mu+\nu}_{\rho}$. As the {\em fiducial} function $\gamma(\rho)$ we take $M_{\alpha}(\sigma)$ for which $c_{\rho}^{n} M_{\alpha}(-\alpha \rho)$ at $\rho = 0$ is equal to $M_{\alpha}(-\alpha n)$. Inserting this observation into Eq. \eqref{2.9} the ML function can be presented as
\begin{equation}\label{2.10}
E_{\alpha}(x) = \E^{x {c}_{\rho}} M_{\alpha}(-\alpha \rho)\big\vert_{\rho=0}.
\end{equation}

\bigskip
%\noindent
{\em The justification of umbral representation of ML function.} \\

Let us now check if Eq. \eqref{2.10} reconstructs the Laplace transform of ML function. To realize this purpose we insert Eq. \eqref{2.10} for $\alpha=l/k$, $l<k$, into the Laplace transform of ML function:
\begin{multline}\label{2.11}
\int_{0}^{\infty} \E^{-x} E_{l/k}(-a x^{l/k}) \D x \\ = \int_{0}^{\infty} \E^{-x - a x^{l/k}c_{\rho}} \D x\, M_{\alpha}(-\alpha\rho)\big\vert_{\rho=0}.
\end{multline}
The second integral in Eq. \eqref{2.11} can be calculated by using \cite[Eq. (2.3.2.13)]{APPrudnikov-v1} according to which Eq. \eqref{2.11} is written as a finite sum of generalized hypergeometric functions. Expressing  this hypergeometric functions in the series form \cite[Eq. (7.2.3.1)]{APPrudnikov-v3}, using the Euler's reflection formula for gamma functions and the splitting formula for sum we get the conventional form of Laplace transform of $E_{l/k}(-a x^{l/k})$, i.e.
\begin{align*}
\begin{split}
\text{LHS of \eqref{2.11}} & = \sum_{r=0}^{\infty}\sum_{j=0}^{k-1} (-a)^{j+kr} \frac{\Gamma\left[1 + \frac{l}{k}(j+kr)\right]}{\Gamma(1+j+kr)} \\
& \times c_{\rho}^{j + kr} M_{l/k}(-l\rho/k)\big\vert_{\rho=0}  \\
&= \sum_{n=0}^{\infty} (-1)^{n} a^{n} = \frac{1}{1+a},
\end{split}
\end{align*}
for $|a| < 1$.

Next, we check if Eq. \eqref{2.10} satisfies the differential equation \cite[Eq. (10.9)]{HJHaubold11}. For that reason we use the relation between the fractional derivative in Riemann-Liouville and the Caputo senses \cite[Eq. (2.134)]{IPodlubny99}. This gives that the fractional Riemann-Liouville derivative of $E_{\alpha}(b x^{\alpha})$, $b\in\mathbb{R}$, reads
\begin{align}\label{2.12}
\begin{split}
& \partial^{\alpha}_{x} E_{\alpha}(b x^{\alpha}) = \frac{x^{-\alpha}}{\Gamma(1-\alpha)} \\
& \quad + \int_{0}^{x}\frac{\alpha b y^{\alpha-1} c_{\rho}\E^{b y^{\alpha} c_{\rho}}}{(x-y)^{\alpha}} \frac{\D y}{\Gamma(1-\alpha)}\,M_{\alpha}(-\alpha \rho)\big\vert_{\rho=0}.
\end{split}
\end{align}
The integral in Eq. \eqref{2.12} can be calculated by employing the series form of exponential function being in the nominator of integrand and taking \cite[Eq. (2.2.4.8)]{APPrudnikov-v1}. Thus, we obtain \cite[Eq. (10.9)]{HJHaubold11} for $0 < \alpha < 1$, this is
\begin{equation}\label{2.13}
\partial^{\alpha}_{x} E_{\alpha}(b x^{\alpha}) = b E_{\alpha}(b x^{\alpha}) + \frac{x^{-\alpha}}{\Gamma(1-\alpha)}.
\end{equation}
The proof that Eq. \eqref{2.10} fulfils Eq. \eqref{2.13} can also be obtained by employing \cite[Eq. (2.3.1.1)]{APPrudnikov-v1} and series representation of the generalised hypergeometric function. Because this way needs tedious calculations it is omitted in the paper.  \\

%\noindent
{\em The shift (umbral) representation of $n_{\alpha}(x, y)$.} \\

We start with taking Eq. \eqref{2.6} in which the L\'{e}vy stable distribution is written in the series form \cite[Eq. (4)]{HPollard46}. Then, we apply $r! = 2^{r}\,\Gamma(1+r/2) \Gamma[(1+r)/2]/\sqrt{\pi}$, the Euler's reflection formula, and Eq. \eqref{2.8}. Thus, we have
\begin{align}\label{2.14}
\begin{split}
& n_{\alpha}(x, y) = \frac{1}{\sqrt{\pi}\alpha} \sum_{r=0}^{\infty} \frac{(-1)^{r+1} \big(\frac{x}{2}\big)^{r-1}}{\Gamma\big(\frac{r+1}{2}\big)\, y^{\alpha r}}\, \frac{\Gamma(1+\alpha r) \sin(\pi\alpha r)}{\Gamma\big(1 + \frac{r}{2}\big)} \\
& \quad = \frac{2}{\sqrt{\pi}} \sum_{r=0}^{\infty} \frac{(-1)^{r+1} \big(\frac{x}{2}\big)^{r-1}}{\Gamma\big(\frac{r+1}{2}\big)\, y^{\alpha r}}\sin(\pi r)\, M_{\alpha}(\alpha r) \\
& \quad = \frac{1}{\sqrt{\pi y^{2\alpha}}} \sum_{n=0}^{\infty} \frac{\big(-\frac{x^{2}}{4 y^{2\alpha}}\big)^{n}}{n!} M_{2\alpha}\Big[\!\!-2\alpha\big(-n-\ulamek{1}{2}\big)\Big].
\end{split}
\end{align}
Introducing the shift (umbral) operator $c_{\rho}$ defined below Eq. \eqref{2.9} we can rewrite Eq. \eqref{2.14} as
\begin{equation}\label{2.15}
n_{\alpha}(\xi, x) = n_{1/2}(\xi, x^{2\alpha} c_{\rho})\, M_{2\alpha}(-2\alpha\rho)\Big\vert_{\rho = 0},
\end{equation}
where
\begin{equation}\label{2.16}
n_{1/2}(\xi, \kappa) = \frac{\exp\!\big(\!-\ulamek{\xi}{4\kappa}\big)}{\sqrt{\pi\kappa}}.
\end{equation}
Remark that the essential singularity of L\'{e}vy stable distribution implies the existence of essential singularity of $n_{1/2}(\xi, x^{2\alpha} c_{\rho})$ at $\xi = 0$. That gives that the series representation of Eq. \eqref{2.16} depends on the couture. The correct way of dealing with the choice of the contour can be obtained by taking Eq. \eqref{2.6} where the one-sided L\'{e}vy distribution is presented in \cite{HPollard46}.

\section{Convolution property of the ML function}\label{sec3} %%%%%%%%%%%%%%%%%%%%%%%%%%%%%

The fact that the product of two ML functions can be written as another ML function appears only for $\alpha = 1$. For the remain value of $\alpha$ this relation is violate. For $\alpha\neq 1$ the convolution property of two ML functions has the form \cite{GDattoli17, GDattoli17a}
\begin{equation}\label{3.17}
E_{\alpha}(\lambda x)E_{\alpha}(\lambda y) = \sum_{n=0}^{\infty}\frac{\lambda^{n} {g}_{\alpha}(n;x, y)}{\Gamma(1+\alpha n)}, \end{equation}
which it can be proved by using the series representation of the ML function. The auxiliary functions ${g}_{\alpha}(n; x, y)$ are given by
\begin{equation}\label{3.18}
{g}_{\alpha}(n; x, y) \,\okr\, \sum_{r=0}^{n} \binom{n}{r}_{\!\alpha} x^{n-r} y^{r},
\end{equation}
where
\begin{equation}\label{3.19}
\binom{n}{r}_{\!\alpha} \,\okr\, \frac{\Gamma(1+\alpha n)}{\Gamma(1+\alpha r) \Gamma[1 + \alpha(n-r)]}.
\end{equation}
Observe that Eq. \eqref{3.19} for $\alpha=1$ is the binomial such that Eq. \eqref{3.18} reduces to the binomial identity and, finally, we get that the sum in Eq. \eqref{3.17} is the ML function. From another side the integral representation of ML function \eqref{2.5} leads to
\begin{equation}\label{3.20}
E_{\alpha}(\lambda x) E_{\alpha}(\lambda y) = \int_{0}^{\infty} \E^{\lambda u} \tilde{n}_{\alpha}(u; x^{1/\alpha}, y^{1/\alpha}) \D u
\end{equation}
with
\begin{equation}\label{3.21}
\tilde{n}_{\alpha}(u; x, y) = \int_{0}^{u} n_{\alpha}(\xi, x) n_{\alpha}(u-\xi, y) \D \xi.
\end{equation}
Comparing now the same terms for fixed $\lambda$ in Eq. \eqref{3.17} and Eq. \eqref{3.20}, where we employ the series representation of $\exp(\lambda u)$ and change the order of integration and summation, $g_{\alpha}(n; x, y)$ reads as
\begin{equation}\label{3.22}
g_{\alpha}(n; x, y) = [M_{\alpha}(-\alpha n)]^{-1}\!\int_{0}^{\infty} u^{n} \tilde{n}_{\alpha}(u; x^{1/\alpha}, y^{1/\alpha}) \D u.
\end{equation}
The auxiliary function $g_{\alpha}(n; x, y)$ can be also derived from the shift (umbral) representation of the ML function given by Eq. \eqref{2.10}. According to it, the multiplication of two ML functions can be straightforwardly written as $E_{\alpha}(\lambda x)E_{\alpha}(\lambda y) = \exp[\lambda(x c_{\rho_{1}} + y c_{\rho_{2}})] M_{\alpha}(-\alpha\rho_{1})\big\vert_{\rho_{1} = 0}M_{\alpha}(-\alpha\rho_{2})\big\vert_{{\rho_{2} = 0}}$. Expressing the exponential function in this formula as the series and comparing the obtained results with Eq. \eqref{3.17} we get $g_{n; \alpha}(x, y) = [M_{\alpha}(-\alpha n)]^{-1}(x c_{\rho_{1}} + y \rho_{z_{2}})^{n}\, M_{\alpha}(-\alpha\rho_{1})\big\vert_{\rho_{1} = 0} M_{\alpha}(-\alpha\rho_{2})\big\vert_{\rho_{2} = 0}$. Note that in \cite{GDattoli17} $g_{n; \alpha}(x, y)$ is named $(x\oplus_{\alpha} y)^{n}$ and the substitution it into Eq. \eqref{3.17} yields to $E_{\alpha}(x) E_{\alpha}(y) = E_{\alpha}(x\oplus_{\alpha} y)$. \\

\noindent
{\em Example.} We calculate $\tilde{n}_{1/2}(u; x, y)$ and $g_{1/2}(n; x, y)$ by using Eqs. \eqref{3.21} and \eqref{3.22}, respectively. Inserting Eq. \eqref{2.16} with $\xi = y$ and $\kappa = x^{2}$ into Eq. \eqref{3.20} and using Eqs. (3.322.1) and (3.322.2) in \cite{ISGradshteyn07} we can write
\begin{align}\label{eq1}
\begin{split}
\tilde{n}_{1/2}(u; x, y) & = \frac{\E^{-\frac{u^{2}}{4(x^{2} + y^{2})}} }{\sqrt{\pi(x^{2} + y^{2})}} \left[{\rm erf}\left(\ulamek{uy}{2x\sqrt{x^2+y^2}}\right) \right.\\ 
&+ \left.{\rm erf}\left(\ulamek{ux}{2y\sqrt{x^2+y^2}}\right)\right].
\end{split}
\end{align}
Thereafter, substituting this formula into Eq. \eqref{3.22}, employing \cite[Eq. (2.8.5.6)]{APPrudnikov-v2} and \cite[Eq. (7.2.1.7)]{APPrudnikov-v3} we obtain
\begin{align}\label{eq2}
\begin{split}
& g_{1/2}(n; x, y) = \frac{2\Gamma(1+n/2)}{\sqrt{\pi}\,\Gamma[(1+n)/2]} \\
& \quad \times \left[y x^{n-1}\, {_{2}F_{1}}\left({1, (1-n)/2 \atop 3/2}; -\frac{y^{2}}{x^{2}}\right) \right. \\
&\quad \left.+ x y^{n-1}\, {_{2}F_{1}}\left({1, (1-n)/2 \atop 3/2}; -\frac{x^{2}}{y^{2}}\right)\right].
\end{split}
\end{align}

The functions $\tilde{n}_{\alpha}(u; x, y)$ and $g_{\alpha}(n; x, y)$ for $\alpha\neq 1/2$ such that $0 < \alpha < 1$ can be derived when we insert Eq. \eqref{2.15} into Eqs. \eqref{3.21} and \eqref{3.22} and use Eqs. \eqref{eq1} and \eqref{eq2}.

\section{The solution of fFP equation}\label{sec4} %%%%%%%%%%%%%%%%%%%%%%%%%%%%%%%%

As is shown in \cite{KGorska12} the use of the series representation of the ML function in Eq. \eqref{1.3} for the Gaussian initial condition and $L_{F\!P} =\frac{\D^{\,2}}{\D x^{2}}$ can lead to divergent series. From this reason we employ the integral and shift (umbral) representations of ML function.

The substitution of Eq. \eqref{2.5} into \eqref{1.3} enable us to present $F_{\alpha}(x, t)$ as the integral \cite[Eq. (13)]{KGorska12}, this is
\begin{equation}\label{4.23}
F_{\alpha}(x, t) = \int_{0}^{\infty} n_{\alpha}(y, t) F_{1}(x, y) \D y, \quad F_{1}(x, y) = \E^{y L_{F\!P}} f(x).
\end{equation}
We remind that the initial condition $f(x)$ is a Fourier transformable function. $F_{1}(x, y)$ is the formal solution of standard (ordinary) FP equation whose definition in Eq. \eqref{4.23} comes from Eq. \eqref{1.3} for $\alpha=1$. For $L_{F\!P} = \frac{\D^{\,2}}{\D x^{2}}$ and $\alpha =1$ Eq. \eqref{1.1} is called the heat (diffusion) equation. Its solution written via the Gauss-Weierstrass transform \cite{GDattoli03a, GDattoli04a, THaimo92, KGorska13, DWidder75} reads
\begin{equation}\label{4.24}
F_{1}(x, y) = \E^{y \frac{\D^{\,2}}{\D x^{2}}} f(x) = \int_{-\infty}^{\infty} \frac{1}{2} n_{1/2}(\xi, y) f(x - \xi) \D\xi,
\end{equation}
where $n_{1/2}(\xi, y)$ is given by Eq. \eqref{2.16}.

Inserting Eq. \eqref{4.24} into Eq. \eqref{4.23} and changing the order of integrations we have
\begin{align}\label{4.25}
\begin{split}
F_{\alpha}(x, t) & = \int_{-\infty}^{\infty} \left[\frac{1}{2} \int_{0}^{\infty} n_{\alpha}(y, t)\, n_{1/2}(\xi, y) \D y\right] f(x - \xi) \D\xi \\
& = \frac{1}{2} \int_{-\infty}^{\infty} n_{\alpha/2}(\xi, t) f(x - \xi) \D\xi.
\end{split}
\end{align}
We recall that $n_{\alpha/2}(\xi, t)$ is given by Eq. \eqref{2.6}. The square bracket in Eq. \eqref{4.25} was calculated by using \cite[Eq. (32)]{KAPenson16}. Next, we substitute Eq. \eqref{2.15} into Eq. \eqref{4.25}. Thus, we get the formula alike Eq. \eqref{4.24} in which $y$ is changes onto $t^{\alpha} c_{\rho}$ where $c_{\rho}$ acts on $M_{\alpha}(-\alpha \rho)$. That allows us to express $F_{\alpha}(x, t)$ via $F_{1}(x, \tau)$ in which the second argument $\tau$ is appropriate changed:
\begin{equation}\label{4.26}
F_{\alpha}(x, t) = F_{1}(x, t^{\alpha} c_{\rho})\, M_{\alpha}(-\alpha\rho)\Big\vert_{\rho = 0}.
\end{equation}
Eq. \eqref{4.26} can also be obtained for arbitrary FP operator by taking the shift (umbral) representation of ML function given by Eq. \eqref{2.10} and inserting it into Eq. \eqref{1.3}. From that reason we treat Eq. \eqref{4.26} as a general property of solution of fFP equation.

Eq. \eqref{4.26} simplifies the calculation of $n$the moments of $F_{\alpha}(x, t)$, i.e.  $\langle x^{n}(t)\rangle_{\alpha} \okr \int_{-\infty}^{\infty} x^{n} F_{\alpha}(x, t)$. Due to Eq. \eqref{4.26} $\langle x^{n}\rangle_{\alpha}$ can be obtained from the $n$th moments of $F_{1}(x, t)$ denoted by $\langle x^{n}(\tau)\rangle_{1}$ where $\tau$ is changed into $t^{\alpha} c_{\rho}$:
\begin{equation}\label{4.27}
\langle x^{n}(t)\rangle_{\alpha} = \langle x^{n}(t^{\alpha} c_{\rho})\rangle_{1} \,M_{\alpha}(-\alpha \rho)\Big\vert_{\rho=0}.
\end{equation}

\section{Examples}\label{sec5} %%%%%%%%%%%%%%%%%%%%%%%%%%%%%%%%%%%%%

In this section we will study few examples of $\langle x^{n}(t)\rangle_{\alpha}$ which correspond to $F_{\alpha}(x, t)$ calculated for various operator $L_{F\!P}$. \\

\noindent
{\bf (1)} We start with the heat (diffusion) equation for which the $n$th moment of $F_{1}(x, \tau)$ is equal to%$^{1}$\footnote{$^{1}$ The number of considered example is denoted by the superscript $(n)$ where $n=1, 2, 3, 4$ and $n=5$.}
\begin{equation}\label{5.28}
\langle x^{n}(\tau)\rangle^{(1)}_{1} \okr \int_{-\infty}^{\infty} x^{n} F_{1}(x, \tau) \D x = \int_{-\infty}^{\infty} f(\xi) H_{n}(\xi, \tau) \D\xi.
\end{equation}
The second equality in Eq. \eqref{5.28} can be obtained by using Eq. \eqref{4.24} where instead of $\xi$ we take $x-\xi$, \cite[Eq. (2.3.15.9)]{APPrudnikov-v1}, and the Hermite Kamp\`{e} de F\'{e}ri\'{e}t polynomials (named also the heat polynomials \cite{DWidder75, PCRosenbloom59}) defined as $H_{n}(X, Y)\okr n! \sum_{r=0}^{\lfloor n/2\rfloor} X^{n-2r} Y^{r}/[(n-2r)! r!]$, $n=0, 1, 2, \ldots$. Note that the polynomials $H_{n}(X, Y)$ can also be derived by calculating $\exp(Y\partial_{X}^{2}) X^{n}$. From Eq. \eqref{4.27} we have
\begin{equation}\label{5.29}
\langle x^{n}(t) \rangle^{(1)}_{\alpha} = \int_{-\infty}^{\infty} f(\xi)\, {_{\alpha}H_{n}}(\xi, t^{\alpha}) \D\xi,
\end{equation}
where the fractional heat polynomials presented in \cite[Eq. (8)]{KGorska12} are equal to ${_{\alpha}H_{n}}(X, Y)\okr n! \sum_{r=0}^{\lfloor n/2\rfloor} X^{n-2r} Y^{\alpha r}/[(n-2r)! \Gamma(1+\alpha r)]$. ${_{\alpha}H_{n}}(X, Y)$ can also be obtained as ${_{\alpha}H_{n}}(X, Y) = E_{\alpha}(Y^{\alpha} \partial_{x}^{2})X^{n}$.

We emphasis that $\langle x^{2}(\tau)\rangle_{1}\sim \tau$ characterises the diffusion process and random walk. Moreover, from Eq. \eqref{5.29} we get $\langle x^{2}(t)\rangle_{\alpha} \sim t^{\alpha}$ which it corresponds to the so-called anomalous (sub- and super-) diffusion and continuous random walk or L\'{e}vy flights. The sub-diffusion is for $0 < \alpha < 1$ and the super-diffusion for $\alpha > 1$. For $\alpha = 1$ we have the normal diffusion. \\

\noindent
{\bf (2)} In the next example we take%$^{1}$
\begin{equation*}
L_{F\!P}^{(2)} = q(x) \frac{\D}{\D x} + v(x),
\end{equation*}
where $q(x)$ and $v(x)$ are arbitrary real functions. Due to Eq. \eqref{4.26} $F^{(2)}_{\alpha}(x, t)$ is obtained from $F^{(2)}_{1}(x, \tau)$ which it is equal to%$^{1}$
\begin{equation}\label{5.30}
F_{1}^{(2)}(x, \tau) = \E^{\tau[q(x)\frac{\D}{\D x} + v(x)]} f(x) = g(x, \tau) f(T(x, \tau)).
\end{equation}
The prove of second equality in Eq. \eqref{5.30} under assumptions
\begin{align}\label{5.31}
\begin{split}
\partial_{\tau} T(x, \tau) & = q(T(x, \tau)),  \qquad T(x, 0) = x, \\
\partial_{\tau} g(x, \tau) & = v(T(x, \tau)) g(x, \tau),  \qquad g(x, 0) = 1,
\end{split}
\end{align}
is presented in \cite{PBlasiak05, GDattoli97}. The list of several applications of Eqs. \eqref{5.30} and \eqref{5.31} for some choices of $q(x)$ and $v(x)$ are given in \cite{PBlasiak05}. Here, we have just quoted one example which will be extensively used in this section. \\

\noindent
From \cite[Ex.1]{PBlasiak05} in which $q(x) = x$ and $v(x) = 0$ we obtain $g(x, \tau) = 1$, $T(x, \tau) = x\E^{\tau}$ and the Euler dilation operator $\exp(\tau x \frac{\D}{\D x})$. In this case Eq. \eqref{5.30} leads to
\begin{equation}\label{5.32}
F_{1}^{(2)}(x, \tau) = f(x \E^{\tau}) \quad \text{and} \quad \langle x^{n}(\tau)\rangle^{(2)}_{1} =  \sigma^{n} \E^{-(n+1)\tau},
\end{equation}
where $\sigma^{n} = \int_{-\infty}^{\infty} \xi^{n} f(\xi)\D\xi$. The $n$th moment of $F_{1}(x, \tau)$ is calculated by using $x = \E^{-\tau}\!\xi$ in the definition of $\langle x^{n}(\tau)\rangle^{(2)}_{1}$. Now, employing Eqs. \eqref{4.27} and \eqref{2.10} we get
\begin{equation*}
\langle x^{n}(t)\rangle^{(2)}_{\alpha} = E_{\alpha}(-(n+1)t^{\alpha}) \sigma^{n}.
\end{equation*}

\bigskip
\noindent
{\bf (3)} The third example is for the operator%$^{1}$
\begin{equation*}
L_{F\!P}^{(3)} = a \frac{\D^{\,2}}{\D x^{2}}  - b \frac{\D}{\D x} x, \qquad a, b > 0,
\end{equation*}
which it is used in storage ring physics to model the effect of diffusion and damping of the electron beam \cite{FCiocci00}. The fFP equation with $L^{(3)}_{F\!P}$ is studied in \cite{KGorska12}, where $\langle x^{2}(t) \rangle^{(2)}_{\alpha}$ is derived by employing the integral representation of ML function. To find the solution of standard (ordinary) FP equation with $L^{(3)}_{F\!P}$ we use the Sack identity \cite[Eqs. (I.2.29c)-(I.2.31)]{GDattoli97} or \cite[Eq. (2.13)]{RSack58} which it says
\begin{equation}\label{5.34}
\E^{A + B} = \E^{A(\E^{\lambda} - 1)/\lambda} \E^{B},
\end{equation}
where $[A, B] = -\lambda A$ with $A = a \ulamek{\D^{\,2}}{\D x^{2}}$, $B = -b \ulamek{\D}{\D x} x$ and $\lambda = 2 b$. It allows one to write Eq. \eqref{5.34} as
\begin{align}\label{5.35}
\begin{split}
F_{1}^{(3)}(x, \tau) & = \E^{-\tau b} \exp\left[\ulamek{a}{2 b}(\E^{2\tau b} - 1)\ulamek{\D^{\,2}}{\D x^{2}}\right] \E^{-\tau b x \frac{\D}{\D x}} f(x) \\
& = \int_{-\infty}^{\infty} \frac{1}{2\sqrt{\pi \mu}} \E^{-\frac{(x-\eta\E^{\tau b})^{2}}{4\mu}} f(\eta) \D\eta,
\end{split}
\end{align}
where $\mu = a (\E^{2\tau b} - 1)/(2b)$. In Eq. \eqref{5.35} we applied Eq. \eqref{5.32} and Eq. \eqref{4.24} in which $\xi = x - \eta\E^{\tau b}$ was taken. From the definition of $\langle x^{n}\rangle_{1}$ given by the first equality in Eq. \eqref{5.28} and the procedure similar like in the example {\bf (1)} we have
\begin{equation}\label{5.36}
\langle x^{n}(\tau)\rangle^{(3)}_{1} = \int_{-\infty}^{\infty} f(\eta) H_{n}\left(\eta\E^{\tau b}, \ulamek{a}{2b}(\E^{2\tau b} - 1)\right) \D\eta,
\end{equation}
where $H_{n}(X, Y)$ is the heat polynomials. Then, due to Eq. \eqref{4.27} the change $\tau$ into $t^{\alpha} c_{\rho}$ in Eq. \eqref{5.36} leads to
\begin{align}\label{5.37}
\begin{split}
& \langle x^{n}(t)\rangle_{\alpha}^{(3)} = n! \sum_{r=0}^{\lfloor n/2\rfloor} \frac{(\frac{a}{2b})^{r} \sigma^{n-2r} }{(n-2r)! r!} \\ 
&\quad \times\E^{n b t^{\alpha} c_{\rho}} (1-\E^{-2 b t^{\alpha} c_{\rho}})^{r}M_{\alpha}(-\alpha \rho)\Big\vert_{\rho=0} \\
&\quad = n! \sum_{r=0}^{\lfloor n/2 \rfloor} \sum_{s=0}^{r} \frac{[a/(2b)]^{r} \sigma^{n-2r}}{s! (r-s)! (n-2r)!} E_{\alpha}(b t^{\alpha} (n-2s)),
\end{split}
\end{align}
where $\sigma^{n}$ is given after Eq. \eqref{5.32}. In Eq. \eqref{5.37} we employed the sum representation of heat polynomials $H_{n}(X, Y)$ and the binomial identity. Note that Eq. \eqref{5.37} for $n=2$ reconstructs the second moment presented in \cite{KGorska12}, i.e. $\langle x^{2}(t)\rangle_{\alpha}^{(3)} = (a/b + \sigma^{2}) E_{\alpha}(2 b t^{\alpha}) - a/b$. \\

\noindent
{\bf (4)} Eqs. \eqref{5.30} and \eqref{5.31} can also be applied in calculation the fFP equation with the operator $L^{(4)}_{F\!P}$ containing the diffusion coefficient being $x^{2}$ and the drift coefficient equals to $x$:%$^{1}$
\begin{equation*}
L_{F\!P}^{(4)} =  \frac{\D}{\D x}\left(x^{2}\frac{\D}{\D x}\right) - \frac{\D}{\D x}x = \left(x \frac{\D}{\D x}\right)^{\!2} - 1.
\end{equation*}
The knowledge of $F_{1}^{(4)}(x, \tau)$ enable one to calculate $\langle x^{n}(\tau)\rangle^{(4)}_{1}$ which after changing $\tau$ onto $t^\alpha c_{\rho}$, where $c_{\rho}$ acts on $M_{\alpha}(-\alpha\rho)$, gives $\langle x^{n}(t)\rangle^{(4)}_{\alpha}$. To get the explicit form of $F^{(4)}_{1}(x, \tau)$ we employ its definition, the integral representation of Gaussian \cite[Eq. (2.3.15.11)]{APPrudnikov-v1}, and the example {\bf (2)}. That allows one to obtain
\begin{align}\label{5.38}
\begin{split}
F_{1}^{(4)}(x, \tau) & = \E^{-\tau}\E^{\tau (x \frac{\D}{\D x})^{2}}f(x) \\
& = \E^{-\tau}\int_{-\infty}^{\infty} \E^{-\frac{u^{2}}{4\tau}} f(x\E^{-u}) \frac{\D u}{2\sqrt{\pi\tau}}.
\end{split}
\end{align}
The proof of Eq. \eqref{5.38} is described in Appendix A. After applying Eqs. \eqref{5.32} and \eqref{2.10} the moments $\langle x^{n}(\tau)\rangle^{(4)}_{\alpha}$ reads
\begin{align*}
\begin{split}
\langle x^{n}(\tau)\rangle^{(4)}_{\alpha} &\!=\!\frac{\E^{-\tau}}{2\sqrt{\pi\tau}}\!\!\int_{-\infty}^{\infty}\!\!\E^{-u^{2}{4\tau}}\!\left[\int_{-\infty}^{\infty} x^{n} f(x \E^{-u}) \D x\right] \D u \Big\vert_{\tau=t^{\alpha} c_{\rho}} \\
& = \frac{\E^{-\tau}}{2\sqrt{\pi\tau}} \int_{-\infty}^{\infty}\E^{-\frac{u^{2}}{4\tau}} \E^{(1+n)u} \D u \Big\vert_{\tau = t^{\alpha} c_{\rho}} \sigma^{n}\\
&= \sigma^{n} \E^{(2n + n^{2}) \tau} \Big\vert_{\tau = t^{\alpha} c_{\rho}} \sigma^{n} = \sigma^{n} E_{\alpha}\left((2n + n^{2})t^{\alpha}\right). \\
\end{split}
\end{align*}

\bigskip
\noindent
{\bf (5)} In the last considered here example we will solve the fFP equation with the FP operator%$^{1}$
\begin{equation}\label{5.39}
L_{F\!P}^{(5)} = \frac{1}{2} \frac{\D^{\,2}}{\D x^{2}} - \frac{1}{2} x^{2},
\end{equation}
which it is intensive used in quantum mechanics, i.e. multiplying it by minus we obtain the hamiltonian of the harmonic oscillator in which mass, frequency, and $\hbar$ are equal to one. Moreover, for complex $x$ the operator \eqref{5.39} correspondences to the $SU(1, 1)$ group \cite{STAli14, GDattoli88} whose generators $S_{\pm}$ and $S_{0}$ satisfy the commutation relation $[S_{-}, S_{+}] = 2 S_{0}$ and $[S_{0}, S_{\pm}] = \pm S_{\pm}$. In the case of $L_{F\!P}^{(5)}$ they have the form $S_{-} = \frac{1}{2} \frac{\D^{\,2}}{\D x^{2}}$, $S_{+} = \frac{1}{2} x^{2}$, and $S_{0} =  \frac{1}{2}(x\frac{\D}{\D x} + \frac{1}{2})$ \cite{STAli14}.  We remark here that the $SU(1, 1)$ group is used to generate the so-called squeezed states. For that reason operator \eqref{5.39} is named the squeezed operator.

Due to Eq. \eqref{4.26} to find $F^{(5)}_{\alpha}(x, t)$ we should know $F^{(5)}_{1}(x, \tau)$. $F^{(5)}_{1}(x, \tau)$ can be calculated by applying the disentanglement formula for squeezed operator, see for instance \cite[Eq. (5.6) for $\xi = -\tau$]{STAli14}, Eqs. \eqref{4.24} and \eqref{5.32}. That leads to
\begin{align*}
\begin{split}
F^{(5)}_{1}(x, \tau) & = \E^{-\frac{T}{2} x^{2}} \E^{\frac{1}{2}{\rm ln}(1-T^{2})(x\frac{\D}{\D x} + \frac{1}{2})} \E^{\frac{T}{2} \frac{\D^{\,2}}{\D x^{2}}} f(x) \\
& = \frac{(1-T^{2})^{1/4}}{\sqrt{2\pi T}} \int_{-\infty}^{\infty} \E^{-\frac{x^{2}}{2 T} + \frac{\sqrt{1-T^{2}}}{T} \xi x} \E^{-\frac{\xi^{2}}{2 T}} f(\xi) \D\xi,
\end{split}
\end{align*}
where $T = \tanh(\tau)$. The knowledge of $F^{(5)}_{1}(x, \tau)$ enables us to derive $\langle x^{n}(\tau)\rangle^{(5)}_{1}$ for its definition which after employing \cite[Eq. (2.3.15.9)]{APPrudnikov-v1} and the definition of heat polynomials $H_{n}(X, Y)$ gets the form
\begin{align*}
\langle x^{n}(\tau)\rangle^{(5)}_{1} & = (1-T^{2})^{1/4} \int_{-\infty}^{\infty} f(\xi) \E^{-\frac{\xi^{2}}{2} T} \\ 
& \times H_{n}(\sqrt{1-T^{2}} \xi, T/2) \D\xi.
\end{align*}

Because the tangent hyperbolic function can be approximate by its argument and cosine hyperbolic function by one we can estimate $T$ and $1-T^{2}$ as $\tau$ and one, respectively. Thus, $F^{(5)}_{1}(x, \tau)$ is alike Eq. \eqref{4.24} where instead of $n_{1/2}(\xi, y)$ we have $n_{1/2}(\xi^{2}, \tau/2)$. The introduce the shift (umbral) operator $c_{\rho}$ leads to $F^{(5)}_{\alpha}(x, t)$ in the form of Eq. \eqref{4.25} with $n_{\alpha/2}(\xi, t)$ is equal to $n_{\alpha/2}(\xi^{2}, 2^{-1/\alpha} t)$. Under this consideration the $n$th moments $\langle x^{n}(\tau)\rangle^{(5)}_{1}$ is proportional to
\begin{align*}
\begin{split}
\langle x^{n}(\tau)\rangle^{(5)}_{1} \propto \int_{-\infty}^{\infty} f(\xi) \E^{-\frac{\xi^{2}}{2}\tau} H_{n}(\xi, \tau/2) \D\xi.
\end{split}
\end{align*}
The use of Eq. \eqref{4.27} and the definition of heat polynomials $H_{n}(X, Y)$ given below Eq. \eqref{5.28} enable us to present $\langle x^{n}(t)\rangle_{\alpha}$ as
\begin{align*}
\begin{split}
& \langle x^{n}(t)\rangle^{(5)}_{\alpha} \propto n! \sum_{r=0}^{\lfloor n/2\rfloor} \frac{(t^{\alpha}/2)^{r}}{r! (n-2r)!} \int_{-\infty}^{\infty} f(\xi) \xi^{n-2r} \\
& \quad \times c_{\rho}^{r} \E^{-\frac{\xi^{2} t^{\alpha}}{2} c_{\rho}} M_{\alpha}(-\alpha \rho)\Big\vert_{\rho=0} \\
& \quad = n! \sum_{r=0}^{\lfloor n/2\rfloor} \frac{(t^{\alpha}/2)^{r}}{(n-2r)!} \int_{-\infty}^{\infty} f(\xi) \xi^{n-2r} E^{1+r}_{\alpha, 1+\alpha r}(-\ulamek{\xi^{2} t^{\alpha}}{2}).
\end{split}
\end{align*}
Here, we take
\begin{align*}
\begin{split}
E^{1+\delta}_{\alpha, 1 + \alpha\delta}(u) & \okr \frac{c_{\rho}^{\,\delta} \E^{u c_{\rho}} }{\Gamma(1+\delta)} M_{\alpha}(-\alpha \rho)\Big\vert_{c_{\rho} = 0} \\ & = \sum_{r=0}^{\infty} \frac{u^{r} c_{\rho}^{r+\delta} }{r! \Gamma(1+\delta)} M_{\alpha}(-\alpha \rho)\Big\vert_{c_{\rho} = 0} \\
& = \sum_{r=0}^{\infty} \frac{(1+\delta)_{r}\, u^{r}}{r! \Gamma(1+\alpha\delta + \alpha r)},
\end{split}
\end{align*}
which it gives the definition of three parameters ML functions \cite{HJHaubold11, FMainardi15, ECapelasdeOliveira11}.

\section{Conclusion}\label{sec6} %%%%%%%%%%%%%%%%%%%%%%%%%

In the paper we consider the formal solution of (1+1)-dimensional fFP equation with the Riemann-Liouville time $t$ ($t \geq 0$) derivative of order $\alpha$ ($0 < \alpha < 1$) and the FP operator which it depends only on position $x$ ($x\in\mathbb{R}$) and its derivative. The standard approach of solving such type of equations bases on applying the integral transform, usually the Laplace and/or Fourier transforms. The use of these techniques allows one to express the solution of fFP equation, $F_{\alpha}(x, t)$, as the Fourier transform of the ML function multiplied by the initial condition $f(x) = F_{\alpha}(x, 0)$, see Eq. \eqref{1.3}. To calculate this Fourier transform we write the ML function in three forms, namely the series, the integral and the shift (umbral) forms. Because the series representation of the ML function can lead to the divergent solution \cite{KGorska12} we apply its shift (umbral) representation which it gives the results equivalent to the results obtained by employing the integral representation of the ML function. That enable us to present $F_{\alpha}(x, t)$ as the solution of the standard (ordinary) Fokker-Planck equation $F_{1}(x, \tau)$ where the time $\tau$ is changed onto $t^{\alpha} c_{\rho}$ with $c_{\rho}$ being the shift (umbral) operator acting on the fiducial states. In consequence, we can expect that some properties of $F_{1}(x, \tau)$ will be visible in the behaviour of $F_{\alpha}(x, t)$. One of such properties is conservation of the norm if the FP operator contains the diffusion term, see examples (1), (3)-(5). Moreover, the umbral (operational) methods allows one to significantly simplify calculations, especially the second moments of $F_{\alpha}(x, t)$ which they are appropriate for the anomalous diffusion in the considered cases.

In the physical and mathematical literatures it can also be found another way of solving Eq. \eqref{1.1}. If the FP operator is proportional to the first derivative over $x$ taken with minus sign we have $\partial_{x} F_{\alpha}(x, t) = - \partial_{t}^{\alpha} F_{\alpha}(x, t) + \frac{t^{-\alpha}}{\Gamma(1-\alpha)}f(x)$, $F_{\alpha}(x, 0) = f(x)$. Its solution is given via the one-sided L\'{e}vy stable distribution and the boundary condition $F_{\alpha}(0, t) = h(t)$, see Eq. (5.2) of \cite{KGorska17} which reads $F_{\alpha}(x, t) = \int_{0}^{\infty} \varPhi_{\alpha}(\xi/x) h(t- \xi) \D\xi/ x$. The use of the method presented in the paper leads to the solution depended on the ML function and $f(x)$. Employing the integral representation of the ML function we can expressed $F_{\alpha}(x, t)$ as Eq. \eqref{4.23} with $F_{1}(x, y) = f(x-y)$. After changing  $y$ onto $(tx/\xi)^{\alpha}$ Eq. \eqref{4.23} can be rewritten as $F_{\alpha}(x, t) = \int_{0}^{\infty} \varPhi_{\alpha}(\xi/x) f(x- (tx/\xi)^{\alpha}) \D\xi/ x$ which has the same integral kernel as in the first approaches shown in this paragraph. The difference is only that in the first view of solving fFP equation we have the boundary and not the initial  conditions.

\section*{Acknowledgments}

The authors thanks Prof. Andrzej Horzela and Dr. Silvia Liccardi for helpful discussion and suggestions.

K.G and A.L. were supported by the NCN, OPUS-12, Program No. UMO-2016/23/B/ST3/01714. 
Moreover, K.G. thanks for support from MNiSW (Poland), Iuventus Plus 2015-2016, Program No. IP2014 013073. % \\

\appendix %%%%%%%%%%%%%%%%%%%%%%%%%%%%%%%%%%

\section{Proof of Eq. \eqref{5.38}}

Let us recall
\begin{equation}\label{1/10-3}
\E^{\lambda \kappa^{2}} = \int_{-\infty}^{\infty} \E^{-\frac{u^{2}}{4\lambda}} \E^{-\kappa u} \frac{\D u}{2\sqrt{\pi\lambda}},
\end{equation}
in which we take $\lambda = \tau$ and $\kappa = x \frac{\D}{\D x}$. According to the operational method Eq. \eqref{1/10-3} with such chosen $\lambda$ and $\kappa$ reads \cite{GDattoli08b, GDattoli06b}
\begin{equation}\label{1/10-4}
F_{1}(x, \tau) = \E^{\tau(x\frac{\D}{\D x})^{2}}f(x) = \int_{-\infty}^{\infty} \E^{-\frac{u^{2}}{4\tau}} \E^{- u x \frac{\D}{\D x}}f(x) \frac{\D u}{2\sqrt{\pi\lambda}}.
\end{equation}
The use of example {\bf(2)} enable us to obtain Eq. \eqref{1/10-4} as the left hand site of Eq. \eqref{5.38} times $\E^{\tau}$.

Observe that the solution \eqref{1/10-4} satisfy the partial differential equation in the form%$^{2}$\footnote{$^{2}$ Remark that $(x\frac{\D}{\D x})^{2}$ is equal to $x^{2}\frac{\D^{\,2}}{\D x^{2}} + x \frac{\D}{\D x}$.}
\begin{equation*}
\frac{\partial}{\partial \tau} F_{1}(x, \tau) = \left(x^{2} \frac{\D^{\,2}}{\D x^{2}} + x \frac{\D}{\D x}\right) F_{1}(x, \tau).
\end{equation*}

 %%%%%%%%%%%%%%%%%%%%%%%%%

\end{document}